\documentclass[prl,aps,showpacs,twocolumn,floatfix]{revtex4}
\usepackage{epsfig} \usepackage{graphics} \usepackage{bm}
\usepackage{amssymb}
\usepackage{graphicx}
\begin{document}

\preprint{Lebed-Wu-PRL}

\title{Phenomenological Approach to the Possible Existence of a Triplet Superconducting Phase in the Quasi-One-Dimensional Conductor Li$_{0.9}$Mo${_6}$O$_{17}$}

\author{A.G. Lebed$^*$}
\author{O. Sepper}

\affiliation{Department of Physics, University of Arizona, 1118 E.
4-th Street, Tucson, AZ 85721, USA}

\begin{abstract}
We consider a theoretical problem of the upper critical magnetic
field parallel to a conducting axis of a quasi-one-dimensional
layered superconductor. We show that the orbital effects against
superconductivity in a magnetic field are capable of destroying the
superconducting phase at low temperatures if the interplane distance
is less than the corresponding coherence length. Applications of our
results to the recent experiments, performed in the superconductor
Li$_{0.9}$Mo$_6$O$_{17}$ [J.-F. Mercure et al., Phys. Rev. Lett.
{\bf 108}, 187003 (2012)], provide strong arguments in favor of a
triplet superconducting pairing in this quasi-one-dimensional
layered conductor.
\end{abstract}


\maketitle

Quasi-one-dimensional (Q1D) superconductors have been intensively
studied since the discovery of superconductivity in the organic
superconductor (TMTSF)$_2$PF$_6$ [1,2]. Early experiments [2-5],
performed on the Q1D superconductors (TMTSF)$_2$X (X=PF$_6$ and
ClO$_4$), provided hints of their unconventional nature. In addition
to possible triplet pairing [6-10], it was proposed that Q1D
superconductors can demonstrate such unusual phenomena as the
reentrant superconductivity [7-10], Larkin-Ovchinnikov-Fulde-Ferrell
(LOFF) phase [2,7-15], and hidden reentrant superconductivity [15].
Although triplet superconducting pairing was considered as the most
probable mechanism for many years, more recent experiments [12,16]
and theories [15,17] are in favor of a $d$-wave like superconducting
pairing in the (TMTSF)$_2$ClO$_4$ conductor. As for
(TMTSF)$_2$PF$_6$ conductor, the question of possible triplet
superconducting pairing [6,7,10,18-20] is still not completely
resolved. Triplet superconductivity is a rather unusual phenomenon.
In our opinion, it has been firmly established in the heavy fermion
superconductor UPt$_3$ [21] and most likely exists in the
Sr$_2$RuO$_4$ [22,23] and ferromagnetic superconductors [24].

In this intriguing situation, it is important that a novel, strong
candidate for triplet superconductivity - the Q1D layered
superconductor Li$_{0.9}$Mo$_6$O$_{17}$ - has been very recently
suggested [25]. As shown in experiments [25], the upper critical
magnetic field, parallel to a conducting axis of this
superconductor, is five times larger than the so-called
Clogston-Chandrasekhar paramagnetic limit for singlet
superconductivity [26]. A distinctive feature of the measurements in
Ref.[25] is that the paramagnetic limit for the superconductor
Li$_{0.9}$Mo$_6$O$_{17}$, $H_p=3.1 \ T$, has been extracted from
direct measurements of the Pauli susceptibility and specific heat
jump (see also Refs.[27,28]). Therefore, it does not depend on the
details of a theoretical model. As also shown in Ref. [25], the
above-mentioned superconductor is in the clean limit and, thus, the
spin-orbital scattering cannot be responsible for the extremely
large experimental value of the upper critical magnetic field along
conducting axis, $H^x_{c2} \simeq 15 \ T$. These facts would posit
strong arguments in favor of the existence of a triplet superconducting
phase in the Li$_{0.9}$Mo$_6$O$_{17}$, making it insensitive to decoupling
Pauli paramagnetic effects.  However, as mentioned in Ref.[25], orbital
destructive effects are minimized when the magnetic field is
parallel to a conducting axis of a Q1D superconductor. Therefore, it
is stressed in Ref.[25] that the experimentally observed destruction
of superconductivity at $H > H^x_{c2} \simeq 15 \ T$ can only be
ascribed to Pauli paramagnetic effects, which is not in favor of the
above mentioned triplet scenario of superconductivity.

The goal of our Letter is to show theoretically that the orbital
effects can destroy superconductivity even for a magnetic field,
applied parallel to a conducting axis of a Q1D layered
superconductor, provided that the inter-plane distance is less than
the corresponding coherence length. By extracting electronic band
and superconducting phase parameters from measurements in Ref.[25],
we show that the above condition holds for the Q1D layered
superconductor Li$_{0.9}$Mo$_6$O$_{17}$. We consider our results as
the second (after Ref.[25]) major step in establishing triplet
superconductivity in the above discussed superconductor.

Let us consider a tight-binding model for the electron spectrum of a Q1D
layered conductor,
\begin{equation}
\epsilon({\bf p})= - 2t_x \cos(p_x a_x) - 2 t_y \cos(p_y a_y) - 2t_z
\cos (p_z a_z),
\end{equation}
where $t_x \gg t_y \gg t_z$ are overlap integrals of the electron
wave functions along ${\bf x}$, ${\bf y}$, and ${\bf z}$
crystallographic axes, respectively. In a magnetic field,
\begin{equation}
{\bf H} = (H,0,0), \ \ {\bf A} = (0,0,Hy),
\end{equation}
parallel to the conducting chains of the Q1D layered conductor (1), it is
convenient to write electron wave function with definite energy and
momentum component $p_x$ in the following way:
\begin{equation}
\psi_{\epsilon, p_x}^{\pm}(x,y,z) = \exp(\pm ip_x x) \exp[\pm
ip^{\pm}_y(p_x)y] \ \phi_{\epsilon,p_x}^{\pm}(y,z) .
\end{equation}
Note that, in Eq.(3), +(-) corresponds to the left (right) sheet of the
Q1D Fermi surface (FS), and the functions $p_y^{\pm}(p_x)$ are
defined by the following equations:
\begin{equation}
v_F(p_x \mp p_F) \mp 2 t_y \cos[p^{\pm}_y(p_x)a_y] =0,
\end{equation}
where $v_F$ and $p_F$ are the Fermi velocity and Fermi momentum,
respectively. In this case, we can linearize Eq. (1) near two sheets
of a Q1D FS in the following way:
\begin{eqnarray}
\delta \epsilon^{\pm}({\bf p})= \pm 2 t_y a_y [p_y-p_y^{\pm}(p_x)]
\sin[p^{\pm}_y(p_x) a_y] \nonumber\\
 - 2t_z \cos (p_z a_z) ,
\end{eqnarray}
where electron energy, $\delta \epsilon = \epsilon - \epsilon_F$, is
counted from the Fermi energy $\epsilon_F$.

In a magnetic field, we use the Peierls substitution method for
Eq.(5), $p_y-p^{\pm}_y(p_x) \rightarrow -i d/dy, \ p_za_z
\rightarrow p_za_z - \omega_z y /v_F$, where $\omega_z=eH v_F
a_z/c$, $e$ is the electron charge, and $c$ is the velocity of
light. As a result, we obtain the following Schr\"{o}dinger-like
equation for the electron wave functions in the mixed $(y,p_z)$
representations:
\begin{eqnarray}
\biggl\{ \mp i v_y[p^{\pm}_y(p_x)] \frac{d}{dy} - 2t_z \cos
\biggl(p_za_z-\frac{\omega_z}{v_F}y \biggl) - 2 \mu_B s H \biggl\}
\nonumber\\
\times \phi_{\epsilon, p_x}^{\pm}(y,p_z) = \delta \epsilon \
\phi_{\epsilon, p_x}^{\pm}(y,p_z),
\end{eqnarray}
with $s$ being the projection of an electron spin on ${\bf x}$ axis;
$\mu_B$ is the Bohr magneton, $v_y[p^{\pm}_y(p_x)]=2t_y a_y
\sin[p^{\pm}_y(p_x)a_y]$. Note that Eq.(6) can be solved exactly:
\begin{eqnarray}
\phi_{\epsilon, p_x}^{\pm}(y,p_z) =  \exp \biggl\{ \frac{\pm i
\delta \epsilon y}{v_y [p^{\pm}_y(p_x)]} \bigg\} \exp
\biggl\{\frac{\pm 2 i \mu_B s H y}{v_y[p^{\pm}_y(p_x)]} \biggl\}
\nonumber\\
\times \exp  \biggl\{ \pm i \frac{2t_z}{v_y[p^{\pm}_y (p_x)]} \int^y_0 \cos
\biggl( p_z a_z - \frac{\omega_z}{v_F} u \biggl) du \biggl\}.
\end{eqnarray}
It is important that the finite temperatures Green functions for the
wave functions (7),(3) can be determined by the standard equation
[29]:
\begin{eqnarray}
g_{i \omega_n}^{\pm}(x,x_1;y,y_1;p_z)= \int^{+\infty}_{-\infty}
d(\delta \epsilon) [\psi^{\pm}_{\epsilon,p_x}(x_1,y_1,p_z)]^*
\nonumber\\
\times \psi^{\pm}_{\epsilon,p_x}(x,y,p_z) / (i \omega_n - \delta
\epsilon),
\end{eqnarray}
where $\omega_n$ is the so-called Matsubara frequency.

In this Letter, we consider the simplest triplet scenario of
superconductivity in the Li$_{0.9}$Mo$_6$O$_{17}$, where
superconducting pairing is not sensitive to Pauli paramagnetic
effects [7]:
\begin{equation}
\hat \Delta(p_x,y) = \hat I  \ sgn(p_x) \ \Delta(y) ,
\end{equation}
where $\hat I$ is a unit matrix in spin-space, $sgn(p_x)$ changes
the sign of a triplet superconducting order parameter on two slightly
corrugated sheets of the Q1D FS, $\Delta(y)$ takes into account the orbital
destructive effects against superconductivity in a magnetic field.
It is important that the triplet order parameter (9) corresponds to
a fully gapped Q1D FS (4), which is in qualitative agreement with
the experimentally observed large specific heat jump in
Li$_{0.9}$Mo$_6$O$_{17}$ superconductor (see Ref.[27]). To derive
the gap equation for superconducting order parameter $\Delta(y)$, we
use Gor'kov's equations for unconventional superconductivity
[30-32]. As a result of the calculations, we obtain:
\begin{eqnarray}
\Delta(y) = g  \biggl< \int_{|y-y_1| > \frac{|v_y(p_y)|}{\Omega}}
\frac{2 \pi T dy_1}{v_y(p_y) \sinh \bigl[ \frac{2 \pi T
|y-y_1|}{v_y(p_y)} \bigl]} \ \Delta (y_1)
\nonumber\\
\times J_0 \biggl\{ \frac{8 t_z v_F}{\omega_z v_y(p_y)} \sin \biggl[
\frac{\omega_z (y-y_1)}{2v_F} \bigg] \sin \biggl[ \frac{\omega_z
(y+y_1)}{2v_F} \bigg] \biggl\} \biggl>_{p_y}  ,
\end{eqnarray}
where $<...>_{p_y}$ stands for the averaging procedure over momentum
$p_y$, $g$ is the electron coupling constant, and $\Omega$ is the cutoff
energy.

Note that Eq.(10) is very general. For instance, at high magnetic
fields and/or low temperatures, it describes the exotic reentrant
superconducting phase, introduced for a different direction of the
magnetic field in Ref.[7]. Analysis of Eq.(10) shows that we can
disregard the reentrant superconductivity effects at high enough
temperatures,
\begin{equation}
T \geq T^*(H) \simeq \frac{\omega_z(H)v^0_y}{2 \pi^2 v_F}  ,
\end{equation}
and low enough magnetic fields,
\begin{equation}
\omega_z(H) \ll \frac{8 t_z v_F}{v^0_y} ,
\end{equation}
where $v^0_y = 2 t_y a_y$. It is possible to show that Eq.(10) can be rewritten
under the conditions (11),(12) in the following way:
\begin{eqnarray}
\Delta(y) = g  \biggl< \int_{|y-y_1| > \frac{|v_y(p_y)|}{\Omega}}
\frac{2 \pi T dy_1}{v_y(p_y) \sinh \bigl[ \frac{2 \pi T
|y-y_1|}{v_y(p_y)} \bigl]}
\nonumber\\
\times J_0 \biggl\{ \frac{4 t_z (y-y_1)}{v_y(p_y)}
\sin \biggl[ \frac{\omega_z
(y+y_1)}{2v_F} \bigg] \biggl\} \Delta(y_1) \biggl>_{p_y}  ,
\end{eqnarray}

It is important that Eq.(13) is still rather general. In fact, it
describes both the so-called Lawrence-Doniach (LD) model [33,34] and
anisotropic 3D superconductivity. As mentioned in Ref.[25], the LD
model condition, $\xi_z \ll a_z/\sqrt{2}$, is not obeyed in the superconductor Li$_{0.9}$Mo$_6$O$_{17}$ and, as we show below, it is possible to simplify
Eq.(13) to describe anisotropic 3D superconductivity:
\begin{eqnarray}
\Delta(y) = g  \biggl< \int_{|y-y_1| > \frac{|v_y(p_y)|}{\Omega}}
\frac{2 \pi T dy_1}{v_y(p_y) \sinh \bigl[ \frac{2 \pi T
|y-y_1|}{v_y(p_y)} \bigl]}
\nonumber\\
\times J_0 \biggl[ \frac{2 t_z \omega_z  (y^2-y^2_1)}{v_y(p_y)v_F}
 \biggl] \Delta(y_1) \biggl>_{p_y}.
\end{eqnarray}
In Eq.(14), it is convenient to perform the following transformation of the variable
$y_1$: $y_1-y = z v_y(p_y)/v_F$. As a result, Eq.(14) can be rewritten as
\begin{eqnarray}
\Delta(y) = g  \biggl< \int_{|z| > \frac{v_F}{\Omega}}
\frac{2 \pi T dz}{v_F \sinh \bigl[ \frac{2 \pi T z}{v_F} \bigl]}
\Delta \biggl[ y+\frac{v_y(p_y)}{v_F} z \biggl]
\nonumber\\
\times J_0 \biggl\{ \frac{2t_z \omega_z}{v^2_F} z \biggl[2y
+\frac{v_y(p_y)}{v_F}z \biggl] \biggl\} \biggl>_{p_y}.
\end{eqnarray}
It is possible to show that in the vicinity of superconducting transition temperature,
$(T_c-T) \ll T_c$, Eq.(15) leads to the Ginzburg-Landau (GL) formula for the upper
critical field,
\begin{equation}
H^x_{c2}(T) = \frac{4  \pi^2 c \hbar T^2_c}{7 \zeta(3) e t_y t_z a_y
a_z} \biggl( \frac{T_c-T}{T_c} \biggl),
\end{equation}
where $\zeta(x)$ is the Riemann zeta function.
We note that the GL slopes for the magnetic field applied perpendicular to the
conducting axis of a Q1D layered superconductor were derived in Refs.[35,36].
Using the results of Ref.[36], we can write
\begin{equation}
H^y_{c2}(T)= \frac{4 \sqrt{2} \pi^2 c T^2_c}{7 \zeta(3) e v_F t_z
a_z} \biggl( \frac{T_c -T}{T_c} \biggl),
\end{equation}
\begin{equation}
H^z_{c2}(T)= \frac{4 \sqrt{2} \pi^2 c T^2_c}{7 \zeta(3) e v_F t_y a_y} \biggl(
\frac{T_c -T}{T_c} \biggl),
\end{equation}
where the GL coherence lengths along ${\bf x}$, ${\bf y}$, and ${\bf z}$ crystallographical
axes are
\begin{equation}
\xi^2_x= \frac{7 \zeta(3)v^2_F \hbar^2}{16(\pi T_c)^2}, \ \xi^2_y =
\frac{7 \zeta(3)t^2_y a_y^2}{8(\pi T_c)^2}, \ \xi^2_z = \frac{7
\zeta(3)t^2_z a_z^2}{8(\pi T_c)^2}.
\end{equation}
From Eqs.(16)-(19) and experimental data [25], $H^x_{c2}(0) \simeq 22 \ T$,
$H^y_{c2}(0) \simeq 4 \ T$, $H^z_{c2}(0) \simeq 1 \ T$, it is
possible to estimate the parameters of the Q1D layered electron
spectrum (1),(4),(5) and the coherence lengths (19). They are
summarized in Table 1.
\newline

\begin{tabular}{|c || c | c |c |c|}
    \hline
    $\bf{Li_{0.9}Mo_6O_{17}}$ & $\bf{\hat{x}}$ & $\bf{\hat{y}}$ & $\bf{\hat{z}}$
    \\ \hline\hline
    $a_i(\AA)$ & 5.53 & 12.73 & 9.51  \\ \hline
    $\xi_i(\AA)$ & 426 & 77& 20 \\ \hline
    $t_i(K)$ & ... & 41 & 14 \\ \hline
    $v_i(cm/s)\cdot 10^6 $ & $v_F=5.3$ & 1.4 & 0.25 \\
    \hline
  \end{tabular} \newline\newline

\begin{figure}[t]
\centering
\includegraphics[width=0.5\textwidth]{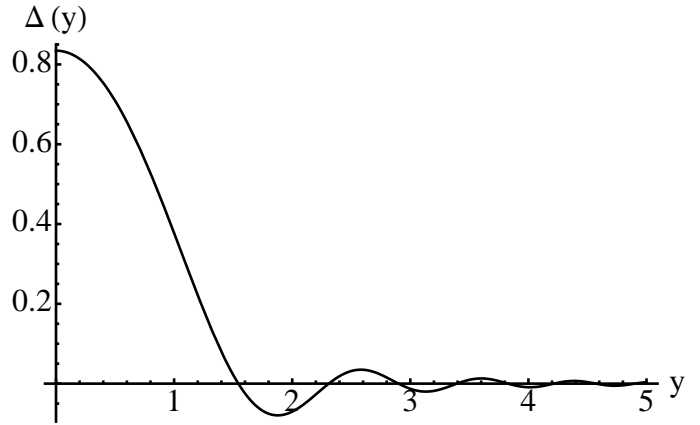}
\caption{Solution of Eq.(15) at $T=0.1 \ K$ is an oscilatory
function of the coordinate y. The solution is normalized by the
following condition: $\int_{-\infty}^{+\infty} \Delta^2(y) dy =1.$}
\end{figure}

By using data from Table 1 and Eqs.(11),(12), we can estimate the
region of temperatures and magnetic fields where Eq.(13) is valid.
As a result, we obtain $T \geq T^* \simeq 0.06 \ K$ and $H \ll 300 \
T$ - conditions, which are well satisfied in experiment [25]. As
already mentioned in Ref.[25] and as seen from Table 1, in the
$\mathrm{Li_{0.9}Mo_6O_{17}}$ superconductor $\xi_z \simeq 20 \AA >
a_z/\sqrt{2} = 6.7 \AA$. Thus, the LD model is not applicable and
we can use Eqs.(14),(15) for anisotropic 3D
superconductivity. We note that Eqs.(14),(15) are qualitatively
different from the gap equations for a 3D isotropic case [37,38],
since the former take into account Q1D topology of the FS (4),(5).
In particular, a typical solution of Eq.(15) at low temperatures, $T
\ll T_c \simeq 2.2 \ K$, changes its sign with changing coordinate
y, in contrast to 3D isotropic case, as shown in Fig.1. We solve
Eq.(15) numerically in the range of temperatures, $T_c \geq T \gg
T^*$, and compare the obtained temperature dependence of the upper
critical magnetic field, $H^x_{c2}(T)$, with the experimental data
[25] in Fig.2. As seen from Eq.(2), the calculated dependence of
$H^x_{c2}(T)$ is in good qualitative and quantitative agreement with
the experimental results.

\begin{figure}[t]
\centering
\includegraphics[width=0.55\textwidth]{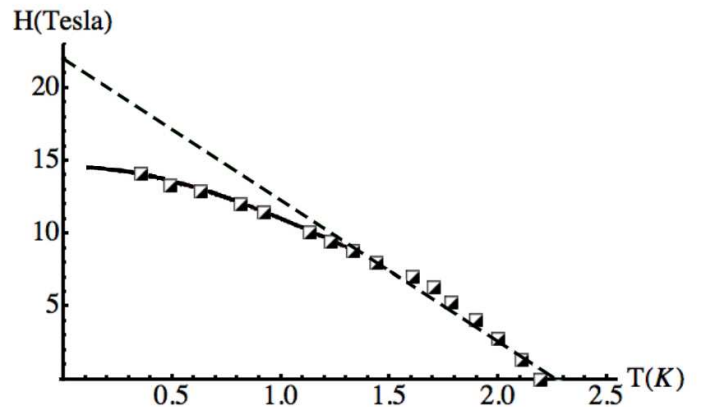}
\caption{Temperature dependence of the upper critical magnetic
field, $H^x_{c2}(T)$, numerically calculated from Eq.(15) al low
enough temperatures, is shown by a solid line. The Ginzburg-Landau
linear dependence (16), which is valid at $T_c-T \ll T_c$, is shown
by broken line. Rectangles represent the experimental data [25]. }
\end{figure}

To summarize, we have explained theoretically the observed
destruction of superconductivity in the Q1D layered superconductor
$\mathrm{Li_{0.9}Mo_6O_{17}}$ in a magnetic field parallel to its
conducting axis [25], in the framework of triplet superconductivity
scenario. We have also suggested the most probable triplet order
parameter [see Eq.(9)]. It corresponds to the absence of Pauli
paramagnetic destructive effects against superconductivity and is
qualitatively consistent with a large value of the experimentally
observed specific heat jump at the superconducting transition at
$H=0$ [27].[Note that from the microscopic point of view, triplet
phase can be stabilized in a Q1D superconductor as a result of
repulsive interchain electron-electron interactions [39-41].]

Below, we would like to discuss the applicability of the Fermi
liquid (FL) approach we have used in the Letter to describe the
superconducting phase transition in the
$\mathrm{Li_{0.9}Mo_6O_{17}}$. First of all, we note that at high
enough temperatures, the Luttinger liquid effects are observed
[42-44] in the $\mathrm{Li_{0.9}Mo_6O_{17}}$ conductor. This
naturally reflects the Q1D nature of its electron spectrum (4),(5)
and is not crucial for our analysis, since we consider the low
temperature region, $T < T_c = 2.2 \ K$. In this context, it is
important that from the theoretical point of view, the FL picture is
restored at temperatures lower than $t_z, t_y \simeq 10-45 \ K$.
Another point of concern is the experimentally observed increase of
resistivity at $T \leq T_{min} \simeq 15-30 \ K$. So far, its nature
has not been clearly understood. At present, there exist two most
popular competing points of view on this resistivity increase
phenomenon: localization effects [45], and the possible partial
charge-density-wave instability or the corresponding fluctuations
[46,47]. Nevertheless, we stress that there are two experimental
features [25], which are important for the validity of our analysis.
The first one is based on the fact that the noted increase in
resistivity is of the order of $\delta \rho / \rho \simeq 0.25$ [25]
and, thus, small. The second feature is that the magnetoresistance
at $T \simeq 4 \ K$, as shown in Ref. [25], demonstrates quadratic
behavior, $\delta \rho (H) / \rho (0) \sim H^2$, which is a direct
test of the FL theory.

One of us (A.G.L.) is thankful to N.N. Bagmet and N.E. Hussey for useful
discussions. This work was supported by the NSF under Grant No
DMR-1104512.

$^*$Also at: L.D. Landau Institute for Theoretical Physics, RAS, 2
Kosygina Street, Moscow 117334, Russia.

\end{document}